\documentclass[conference]{IEEEtran}
\IEEEoverridecommandlockouts
\usepackage{cite}
\usepackage{amsmath,amssymb,amsfonts}
\usepackage{algorithmic}
\usepackage{graphicx}
\usepackage{textcomp}
\usepackage{xcolor}
\usepackage{url}
\def\BibTeX{{\rm B\kern-.05em{\sc i\kern-.025em b}\kern-.08em
    T\kern-.1667em\lower.7ex\hbox{E}\kern-.125emX}}
\begin{document}

\title{ARCHE: Autoregressive Residual Compression with Hyperprior and Excitation\\
}

\author{\IEEEauthorblockN{Sofia Iliopoulou}
\IEEEauthorblockA{\textit{Dept. of Electrical and Computer} \\
\textit{Engineering, University of Patras}\\
Patras, Greece \\
sofia\_iliopoulou@ac.upatras.gr}
\and
\IEEEauthorblockN{Dimitris Ampeliotis}
\IEEEauthorblockA{\textit{Dept. of Digital Media and} \\
\textit{Communication, Ionian University}\\
Argostoli, Greece \\
ampeliotis@ionio.gr}
\and
\IEEEauthorblockN{Athanassios Skodras}
\IEEEauthorblockA{\textit{Dept. of Electrical and Computer} \\
\textit{Engineering, University of Patras}\\
Patras, Greece \\
skodras@upatras.gr}
}

\maketitle

\begin{abstract}
Recent progress in learning-based image compression has demonstrated that end-to-end optimization can substantially outperform traditional codecs by jointly learning compact latent representations and probabilistic entropy models. However, many existing approaches achieve high rate–distortion efficiency at the expense of increased computational cost and limited parallelism. This paper presents ARCHE – Autoregressive Residual Compression with Hyperprior and Excitation, an end-to-end learned image compression framework that balances modeling accuracy and computational efficiency. The proposed architecture unifies hierarchical, spatial, and channel-based priors within a single probabilistic framework, capturing both global and local dependencies in the latent representation of the image, while employing adaptive feature recalibration and residual refinement to enhance latent representation quality. Without relying on recurrent or transformer-based components, ARCHE attains state-of-the-art rate–distortion efficiency: it reduces the BD-Rate by approximately 48\% relative to the commonly used benchmark model of Balle et al., 30\% relative to the channel-wise autoregressive model of Minnen \& Singh and 5\% against the VVC Intra codec on the Kodak benchmark dataset. The framework maintains computational efficiency with 95M parameters and 222ms running time per image. Visual comparisons confirm sharper textures and improved color fidelity, particularly at lower bit rates, demonstrating that accurate entropy modeling can be achieved through efficient convolutional designs suitable for practical deployment.
\end{abstract}


\section{Introduction}
The ever-growing demand for high-quality visual content has made efficient image compression a critical component of modern communication and storage systems. The constant use of digital images in social media, cloud platforms and other applications has made the transmission of compact imagery of high visual quality a necessity. Traditional compression standards such as JPEG \cite{wallace1992jpeg}, JPEG2000 \cite{rabbani2002jpeg2000,skodras2001jpeg2000}, and JPEG XL \cite{alakuijala2019jpegxl} have achieved remarkable success over the past decades, relying on carefully hand-crafted transforms, quantisation rules, and entropy coding schemes. However, their adaptability to the large variability of natural images is limited, which restricts further improvement – especially as images become more complex and high-resolution.

In recent years, learning-based image compression has emerged as a powerful alternative to traditional codecs. Instead of depending on fixed transforms and manually designed bit allocations, these models jointly optimise the entire compression process from feature extraction and quantisation to entropy modelling and reconstruction, utilising neural networks that learn compact and statistically efficient representations of images which match their true distributions better \cite{esenlik2025overview,ghorbel2023convnext,wang2024learned}. By improving the trade-off between compression ratio and reconstruction quality, learning-based methods are constantly pushing the limits of visual data compression.

The success of this type of techniques stems from their formulation as end-to-end variational frameworks, where analysis and synthesis transforms are trained jointly with a learned entropy model \cite{cui2021asymmetric}. These systems can automatically learn structured latent representations that capture both global semantics and fine-grained spatial details, consistently outperforming traditional compression methods. Progress in this field has been driven by several directions: the introduction of hierarchical priors to transmit side information for entropy estimation \cite{zhou2024lightweight}, the use of adaptive transforms that better align latent statistics with image structures, and the experimentation with more expressive probabilistic models capable of approximating complex data distributions \cite{fu2023gaussian} 

Despite these advances, several challenges remain. State-of-the-art models often achieve great rate–distortion performance at the expense of higher computational cost and slower decoding. Frameworks based on attention mechanisms or transformers give excellent visual results but are difficult to deploy efficiently \cite{koyuncu2022contextformer,xu2024window,chen2023transtic}. On the other hand, sequential entropy models restrict parallel processing and exhibit slow inference \cite{ali2023efficient, minnen2018joint}. Therefore, the balance between model expressiveness, parameter efficiency and practical feasibility remains difficult to achieve. As the field grows, emphasis is shifting toward architectures that combine strong compression performance with practical applicability across diverse platforms \cite{rippel2017realtime,zhang2024efficient}.

Several recent studies have attempted to tackle this issue by taking a hybrid approach. Instead of relying on a single global or local modelling architecture, they attempt to capture dependencies across scales, spatial regions, and feature channels while remaining computationally efficient \cite{guo2022causal}. Meanwhile, advances in probabilistic modelling, such as mixture-based entropy models, offer greater flexibility in handling multimodal latent distributions \cite{zhu2022unified,cheng2020learned}. These developments reflect a broader trend toward efficient compression frameworks that incorporate the strengths of different modelling techniques into a cohesive system.

Building upon recent advances in learned image compression, this work introduces an end-to-end framework that integrates several complementary modelling approaches into a compact and efficient architecture. The key components of the proposed model include: (a) autoregressive (AR) modelling for sequential latent representation prediction, (b) masked context adaptation to improve entropy estimation, (c) channel conditioning (CC) along with a squeeze-and-excitation module for better latent distribution estimation, (d) hyperprior-based entropy modelling to capture global dependencies and (e) latent residual prediction to minimise reconstruction errors. The design aims to balance compression performance and computational efficiency, maintaining a fully convolutional, end-to-end trainable structure. Instead of increasing architectural complexity, the framework focuses on enhancing the expressiveness of latent representations and the accuracy of entropy estimation through better modelling of statistical dependencies. 

The main contributions of this work can be summarised as follows:

\begin{itemize}
\item \textbf{Efficient rate-distortion performance:}
Extensive experiments demonstrate that the proposed method delivers competitive rate–distortion performance compared to state-of-the-art frameworks.

\item \textbf{Lower computational cost:}
The method operates with reduced computational overhead, offering faster running and a lighter architecture while maintaining reconstruction quality.

\item \textbf{Improved visual fidelity, especially at low bit rates:}
Across standard benchmarks, the model consistently produces sharper edges, smoother textures, and more natural colour transitions, particularly at low bit rates.

\item \textbf{Transformer-free and recurrence-free design:}
These improvements are achieved without relying on heavy transformer architectures or recurrent modules, highlighting the effectiveness of a carefully designed convolutional architecture for efficient learned image compression.
\end{itemize}

The rest of this paper is structured as follows: Section II reviews prior work on learned image compression, focusing on hierarchical, probabilistic, and hybrid modelling approaches. Section III introduces the proposed framework and its core components. Section IV describes the experimental setup and evaluation methodology, with results presented in Section V. Finally, Section VI summarises the findings and discusses directions for future research.

\section{Related Works}

Over the past few years, learned image compression has evolved rapidly, transitioning from fixed transform algorithms to fully trainable architectures that jointly learn analysis transforms, entropy models, and rate–distortion trade-offs within a unified probabilistic framework or a transformer network. The use of end-to-end neural networks for image compression has grown rapidly over the past decade, with variational autoencoders (VAEs) providing a framework for jointly learning transforms and entropy models. Early work by Ballé et al. demonstrated that a convolutional autoencoder trained under a rate–distortion loss can approximate classical transform coding architectures \cite{balle2017endtoend}. Building on this, Ballé et al. introduced a learned hyperprior model, in which side information communicates the distribution of latent codes to better estimate their entropy \cite{balle2018variational}. The hyperprior model remains a baseline of learned image compression. Bao et al. analyzed nonlinear transforms in learned image compression from a communication theory perspective, providing theoretical insights into the rate-distortion benefits of learned transformations over fixed linear transforms \cite{bao2023nonlinear}. Li et al. developed statistical measurements to evaluate the robustness and the channel importance of latent representations, providing insights into the distributional properties that affect compression efficiency \cite{li2024revisiting}. Although single-Gaussian priors have been effective in these frameworks by assuming zero-mean Gaussian distributions for every code, the latent space often displays spatial dependencies that such simple factorized priors fail to capture. To address this constraint, later works incorporated hierarchical latent models and mixture distributions, improving the match between the learned latent density and the real coefficient statistics \cite{cheng2020learned,zhang2025generalized}.

Later, a significant development in learned image compression with entropy modelling was the use of autoregressive context models to capture spatial dependencies among latent codes. Minnen et al. proposed a combined autoregressive and hyperprior model where latent elements are conditioned on previously decoded neighbours in raster order, improving the accuracy of entropy estimation \cite{minnen2018joint}. Lee et al. applied masked convolution networks based on PixelCNN \cite{oord2016pixel} as context models to further improve local spatial modelling in the latent space \cite{lee2019extended}. On the other hand, Minnen \& Singh proposed a channel-wise autoregressive prior, directing the dependency modelling along the feature‐channel dimension instead of the spatial domain, thus improving parallelism and reducing decoding latency \cite{minnen2020channelwise}. Although full spatial autoregression provides substantial coding improvements, it creates sequential bottlenecks. Hybrid approaches, like Guo et al.’s three-dimensional context model that combines spatial and channel dependencies, present intermediate trade-offs \cite{guo20203d}. Li et al. proposed a multi-rate progressive entropy model that interprets split-coded-then-merge entropy strategies as a filter banks framework, dividing latent features into nonuniform subsets based on spatial and channel sampling rates to enhance both rate-distortion performance and decoding speed \cite{li2024multirate}. Cheng et al. proposed discretized Gaussian mixture likelihoods and incorporated attention mechanisms to increase the flexibility of the conditional likelihoods used in entropy models \cite{cheng2020learned}. Chen et al. integrated non-local attention mechanisms for capturing long-range dependencies in latent representations, showing that global context modelling complements local autoregressive priors \cite{chen2021endtoend}.  Guo et al. proposed a causal contextual prediction model that captures dependencies across scales and spatial regions while preserving computational efficiency through structured masking operations \cite{guo2022causal}.

More recently, transformer architectures and attention mechanisms have been employed in the learned compression field, providing global context modelling beyond local convolutional receptive fields. Liu et al. proposed a hybrid CNN–Transformer codec combining a Swin-Transformer attention block with channel-wise entropy modelling \cite{liu2023mixed}. Zou et al. developed a window-based local attention framework used in both the encoder and the decoder that featured symmetrical transformer blocks to better capture long-range dependencies in the image \cite{zou2022devil}. Meanwhile, Yang et al. combined neighbourhood window attention with enhanced SwinT and CNN blocks to improve both global and local latent modelling \cite{yang2024accurate}. These transformer-based frameworks achieve excellent rate-distortion performance. However, this comes at the cost of increased model size, inference latency and computational complexity. 

Combining transform-domain techniques with learning-based algorithms has demonstrated potential for improved compression and latent decorrelation. Fu et al. worked in the wavelet domain, developing the WeConvene framework which introduced wavelet-domain channel-wise autoregression and convolution modules to improve multi-scale latent sparsity and entropy modelling \cite{fu2024weconvene,fu20253dwm}. Ma et al. proposed an end-to-end optimised codec that employs a wavelet-like transform that generalises traditional multi-resolution decomposition within a trainable neural framework, thus enabling more compressed latent representations and improved scalability across bit rates \cite{ma2020iwave}. Mishra et al. incorporated wavelet transforms into deep autoencoder architectures, showing that combining traditional signal processing techniques with learned components can improve compression efficiency \cite{mishra2021wavelet}. Meanwhile, Meyer et al. introduced an efficient wavelet-based framework for learned image and video coding that replaces conventional convolutional transforms with learned wavelet analysis and synthesis filters, achieving higher parallelism and lower decoding latency without decreasing the rate–distortion performance \cite{meyer2024efficient}. Wu et al. proposed a learned block-based hybrid approach that combines traditional block-wise processing with neural networks, offering a practical compromise between traditional codecs and learning-based systems \cite{wu2022learned}.

An increasing amount of research highlights the significance of inference speed, model size, and decoding inference in learned image compression codecs. Duan et al. examined the balance between encoder and decoder complexity for image compression in classification tasks, emphasizing the trade-off between rate-distortion performance and computational cost \cite{duan2024balancing}. Additionally, the wavelet‐based iWave++ model achieved speedups of over 350 times in image compression by substituting parallelised context models for sequential ones, while having only a small increase in BD‐rate \cite{ma2022versatile}. These results are centered around efficiency and highlight the importance of balancing entropy modelling accuracy with decoding speed, making them particularly relevant for practical applications of learned image compression frameworks. He et al. further addressed this trade-off by introducing ELIC, a learned image compression framework that employs unevenly grouped space–channel contextual adaptive coding to balance modelling accuracy and parallel processing \cite{he2022elic}. Zhang et al. analysed efficient architectural choices and demonstrated that lightweight CNN designs can substantially reduce complexity while maintaining strong rate-distortion performance \cite{zhang2024efficient}. 
\begin{table}[t]
\centering
\caption{Comparison of Entropy Modelling Approaches}
\label{tab:comparison}
\resizebox{\columnwidth}{!}{%
\begin{tabular}{|l|c|c|c|c|}
\hline
\textbf{Method} & \textbf{Hyperprior} & \textbf{AR} & \textbf{CC} & \textbf{Context Type} \\
\hline
Ballé et al. \cite{balle2018variational} & Yes & No & No & None \\
Minnen et al. \cite{minnen2018joint} & Yes & Yes & No & Masked Convolution \\
Minnen \& Singh \cite{minnen2020channelwise} & Yes & No & Yes & None \\
Iliopoulou et al. \cite{iliopoulou2025learned} & Yes & Yes & Yes & LSTM \\
\textbf{ARCHE (proposed)} & \textbf{Yes} & \textbf{Yes} & \textbf{Yes} & \textbf{Masked PixelCNN + SE} \\
\hline
\end{tabular}%
}
\end{table}
In summary, the field of learned image compression is increasingly moving beyond purely convolutional autoencoder
and prior algorithms, toward architectures that incorporate context models, attention modules, transformer blocks and transform‐domain compression, while balancing computational efficiency and decoding latency. The proposed architecture ARCHE builds on this evolution by integrating a hierarchical hyperprior, masked spatial autoregressive context modelling, channel‐wise conditioning and squeeze‐and‐excitation channel recalibration. Unlike methods that only utilise transformers, ARCHE retains convolutional efficiency while achieving strong modelling accuracy. Meanwhile, it also strikes a balance between parallel processing and entropy modelling expressiveness, unlike purely channel-wise or spatially autoregressive models. Table \ref{tab:comparison} summarises key architectural differences between ARCHE and similar entropy modelling approaches.
\begin{figure*}[t]
\centering
\includegraphics[width=\textwidth]{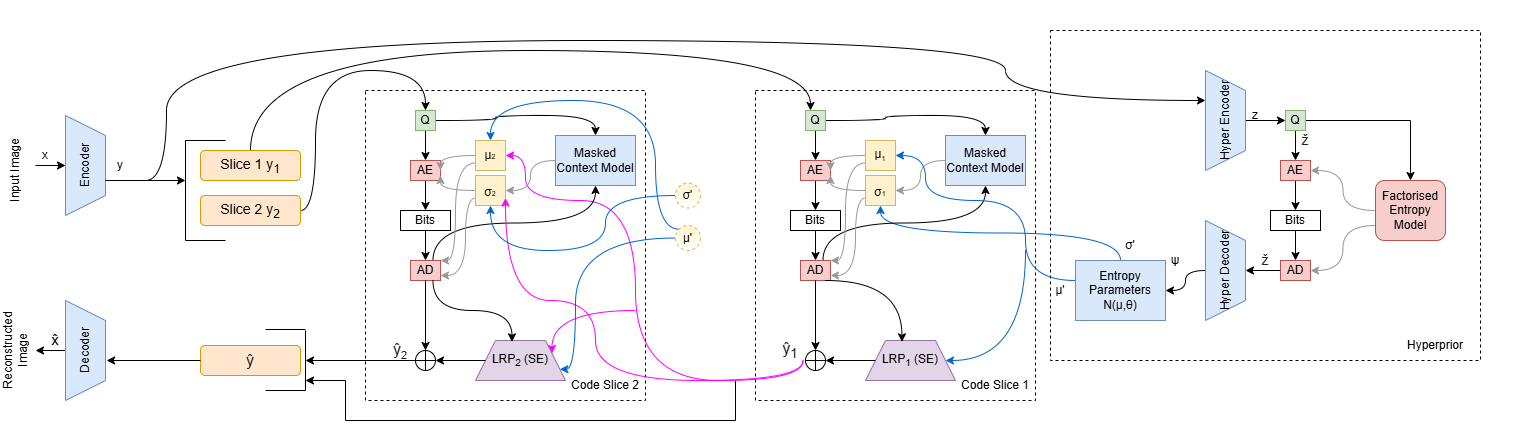}
\caption{ARCHE architecture with hyperprior, masked context model, latent residual prediction, and two example slices showing channel conditioning with SE blocks. Q: quantization; AE/AD: arithmetic encoder/decoder; $\mu$, $\sigma$: distribution parameters.}
\label{fig:architecture}
\end{figure*}

\begin{figure}[b]
\centering
\includegraphics[width=\columnwidth]{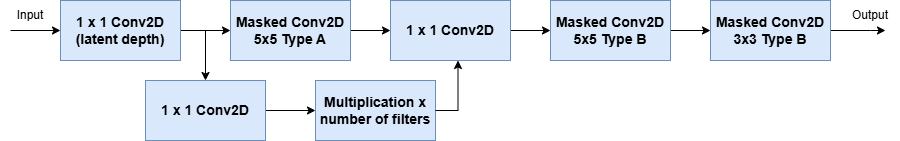}
\caption{The architecture of the masked context model. Type A masked convolution zeroes out all pixels below and to the right of the center (including the center). Type B is the same but includes the center pixel itself.}
\label{fig:maskedcontext}
\end{figure}

\section{The Proposed Method}

Several modern learned image compression methods are structured as hierarchical VAEs, where quantisation and entropy coding are integrated within the latent-variable model \cite{duan2023lossy}. The ARCHE architecture is also based on the form of a variational autoencoder. It consists of an analysis transform $g_a$, a synthesis transform $g_s$, a hyperprior pair ($h_a$, $h_s$), autoregressive residual and excitation modules and Masked PixelCNN-based context modelling. Fig. \ref{fig:architecture} presents a high-level overview of the proposed architecture. Given an input image $x$, the goal is to obtain a compact latent representation $y$ that minimises the rate-distortion cost:
\begin{equation} \label{eq:rate}
L = R + \lambda D
\end{equation}
where $R$ denotes the expected bit rate, $D$ represents some distortion metric (e.g., the mean squared error - MSE), and $\lambda$ regulates the trade-off between them. In end-to-end learned image compression, the latent representation often preserves spatial correlations due to the limited receptive field of convolutional operations. To mitigate this redundancy, entropy modelling is employed within the entropy-constrained bottleneck of the neural network. The purpose of compression extends beyond mere dimensionality reduction—it also aims to minimise the entropy of the latent code using a probabilistic prior shared between the encoder and decoder, known as the \textit{entropy model}. This model estimates the conditional probability distribution of the latent variables given the side information and, when combined with standard lossless entropy coding techniques such as arithmetic coding, generates a more compact bitstream.

 Entropy modelling, which provides an estimate of the code rate, plays a pivotal role in learning-based compression methods. The proposed entropy model adopts a hierarchical design: a global prior is first inferred through the hyperprior, followed by progressive refinement of the probability estimates via channel-conditional and autoregressive context modelling. Finally, latent residual prediction is applied, to mitigate the residual quantisation errors. In the following paragraphs, we explain the constituent parts of the proposed method in greater detail.
 
\subsection{Variational Autoencoder}

Within the transform coding approach for compressing images, an image $x$ is transformed through a parameterized analysis transform $g_a(x;\phi_g)$ into a latent representation $y$, which subsequently undegoes quantization to create a discrete latent vector $\hat{y}$. Given that $\hat{y}$ is discrete, it can undergo lossless compression through entropy coding – such as arithmetic coding – to generate a compact bitstream. The decoder rebuilds $\hat{y}$ from this bitstream and utilises a synthesis transform $g_s(\hat{y};\theta_g)$ to acquire the reconstructed image $\hat{x}$. 

In the proposed variational compression framework, the encoder and decoder are implemented as deep nonlinear transforms parameterized by convolutional neural networks. The analysis transform $g_a(x;\phi_g)$ maps an input image $x$ into a latent representation $y$, which captures spatially compact and statistically independent features suitable for entropy coding. The transform is composed of a sequence of convolutional layers with progressively increasing channel depth and downsampling stride, each followed by Generalized Divisive Normalization (GDN) nonlinearities. GDN has been shown to Gaussianise feature activations and improve the efficiency of density modelling, proving especially effective for learned compression. The analysis transform is formally expressed as $g_a(x;\phi_g)$, where the parameters $\phi_g$ denote the learned convolutional kernels and normalization parameters. The synthesis transform $g_s(\hat{y};\theta_g)$ reconstructs the image from the quantized latent representation $\hat{y}$. It mirrors the analysis path using transposed (upsampling) convolutions combined with Inverse GDN (IGDN) activations to invert the local normalization and recover spatial correlations. The synthesis transform reconstructs the image as $\hat{x}$ = $g_s(\hat{y};\theta_g)$, where $\theta_g$ denote the parameters of the decoder network. 

During training, the quantization is approximated by uniform additive noise, enabling gradients to propagate through the encoder and decoder while avoiding the “vanishing gradients” problem that is often seen in deep networks \cite{balle2016endtoend}. At inference, actual rounding and entropy coding are employed to produce the discrete representation $\hat{y}$, which is then used by the synthesis transform to reconstruct the image. The relaxed model can be interpreted as a variational autoencoder, in which the analysis transform serves as an approximate inference network $q(\hat{y}\mid x)$ and the synthesis transform corresponds to a generative model $p(x \mid \hat{y})$, where $\hat{y}$ represents the noisy latent representation used during training. 

\subsection{Autoregressive Hyperprior}

The latent representation $y$ that is produced by the analysis transform captures compact spatial features, but it generally displays significant local correlation and structured dependencies that restrict compression efficiency when represented by a fully factorized prior. To address this and mitigate the redundancies, the proposed model employs a hierarchical entropy model that combines a global hyperprior with a spatially autoregressive prior, thereby improving the accuracy of the learned probability estimates.

The hyperprior captures global statistical variations in the latent space, particularly the spatially varying uncertainty and scale of the features. The encoder maps the primary latent representation $y$ to a secondary latent variable $z$ via a hyper-analysis transform $h_a(y;\phi_h)$, which captures coarse-scale information such as the magnitude and structure of $y$. The quantized version $\hat{z}=Q(z)$ is transmitted as side information, with its distribution modelled by a learned prior $p_{\hat{z}}(\hat{z})$. As for the decoder, $\hat{z}$ is processed by a hyper-synthesis transform $h_s(\hat{z};\theta_h)$, which reconstructs parameters—typically the predicted means and scales—of the conditional prior $p_{\hat{y}\mid\hat{z}}(\hat{y}\mid\hat{z})$. This mechanism enables adaptive entropy modelling conditioned on the content of the image, allowing the model to allocate bits more efficiently across areas of varying complexity. 

Although the hyperprior takes global variations into account, it assumes that individual latent elements are conditionally independent given $\hat{z}$. In practice, strong spatial dependencies remain among adjacent coefficients of $\hat{y}$, stemming from the limited receptive field of convolutional transforms. To capture these local dependencies, a spatially autoregressive prior is incorporated, parameterized by a sequence of masked convolutional layers. The autoregressive prior models the conditional distribution of each latent element $\hat{y}_i$  based on all previously decoded elements arranged in a specified spatial order:
\begin{equation}
p_{\hat{y}|\hat{z}}(\hat{y}|\hat{z}) = \prod_i p(\hat{y}_i | \hat{y}_{<i}, \hat{z})
\end{equation}
where $\hat{y}_{<i}$ refers to all elements that precede spatially in raster-scan order. Every conditional probability is defined by mean and scale values estimated from both the hyper-synthesis output conditioned on $\hat{z}$ which offers global statistical context, and the autoregressive path implemented through masked convolutions that ensure the causality of the model.

The masked convolutions restrict the receptive field of each prediction so that only previously decoded elements influence the estimation of the current symbol’s parameters. This design allows the model to predict and encode each element sequentially while preserving a valid dependency structure. This spatially autoregressive mechanism refines the conditional entropy estimates beyond the capabilities of a strictly hierarchical model, efficiently utilising local patterns in the latent feature maps. The combination of the hierarchical hyperprior and the spatial autoregressive prior results in a robust joint entropy model that captures both global variations and local spatial correlations, thereby significantly improving the accuracy of the probability estimates used in entropy coding. 

\subsection{Masked Autoregressive Context Model}

Although the hierarchical hyperprior and spatially autoregressive prior address global and sequential dependencies, they fail to completely account for the fine-grained local correlations among latent elements that remain after the analysis transform. To overcome this limitation, the proposed framework incorporates a masked autoregressive context model that explicitly utilises the spatial structure of the quantised latent representation. This context model is based on PixelCNN and adapted to operate in the latent space rather than directly on pixel intensities. This component aims to refine entropy estimates by modelling the conditional distribution of each latent element based on its spatial neighbourhood.

Models derived from PixelCNN provide an effective autoregressive mechanism to model fine-grained local dependencies using convolutions with causal masking, ensuring that each prediction depends only on previous spatial positions. In contrast to recurrent architectures such as Convolutional Long Short-Term Memory Networks (ConvLSTMs), which process the latent tensor sequentially along a scan line \cite{johnston2018improved}, the masked convolution can operate in parallel across spatial dimensions using a single forward pass and still maintain strict autoregressive causality. Consequently, the decoding speed and the computational efficiency are significantly improved.

The main concept is to model the conditional probability of each latent element $\hat{y}_i$ not only as a function of the hyperprior features $\hat{z}$ and previously decoded elements $\hat{y}_{<i}$, but also by extracting context features from its spatial surroundings. The conditional prior is defined as:
\begin{equation}
p(\hat{y}_i|\hat{y}_{<i},\hat{z}) = \mathcal{N}(\hat{y}_i; \mu_i, \sigma_i^2)
\end{equation}
where N is the Gaussian distribution, $\mu$ is the mean value and $\sigma$ is the scale of each element $i$. These parameters are predicted by a parameter estimation network that incorporates information from both the autoregressive hyperprior and the autoregressive context model. 

The context model utilises masked convolutional layers for its implementation. Fig. \ref{fig:maskedcontext} presents the architecture of the context model. For a given position $i$, the receptive field of the masked kernel includes only elements that precede it spatially in raster-scan order (i.e., upper and left neighbours). This constraint is achieved by applying a binary mask to the convolutional filters, setting future positions in the kernel to zero. Consequently, the model is able to effectively estimate the conditional statistics of each latent element during the sequential decoding while maintaining the autoregressive dependency structure.

The masked context model generates a spatial feature tensor that captures local prediction features, such as edge structure, texture consistency, and co-occurrence patterns across adjacent latent elements. These features are lightweight but contain a lot of information, allowing for precise local adaptation of the predicted mean and scale parameters. The outputs of the hyperprior and the masked context model are concatenated and processed by the parameter network, which consists of pointwise and small-kernel convolutions followed by nonlinear activations. This network produces the Gaussian parameters ($\mu$,$\sigma$) which are used by the entropy model to compute the log-likelihood:
\begin{equation}
-log_2p_{\hat{y}|\hat{z}} =  \sum_{i} [-\log_2 \mathcal{N}(\hat{y}_i; \mu_i, \sigma_i^2)]
\end{equation}

To expand the receptive field and capture richer dependencies, multiple masked convolutional layers are stacked, along with nonlinearities such as the sigmoid function. This deep stacking allows the context model to incorporate information from a broader area while retaining causal structure. The final output forms the context feature tensor, which encodes all spatial information available for the current element $i$. The context feature vector is concatenated with the global features vector from the hyper-synthesis transform and fed to the parameter estimation network, which then produces the conditional mean and scale parameters for each latent element’s distribution. The predicted parameters define a Gaussian prior used by the entropy coder.

\subsection{Channel Conditioning}
While the integration of the hyperprior and the masked autoregressive context model successfully captures global and spatial dependencies, its redundant inter-channel correlations remain in the latent representation. The channel features produced by the analysis transform are not statistically independent; different channels encode complementary structural or textural details of the image. In order to take advantage of these residual dependencies, the proposed framework includes a channel conditioning mechanism that adaptively adjusts the entropy parameter estimation for each latent channel, utilising the decoded information from other channels along with the priors.

Let $\hat{y}$ represent the quantised latent tensor that consists of C channels. The entropy model predicts a conditional distribution $p(\hat{y}_{i,c}\mid \hat{y}_{<i,c},\hat{y}_{<c}, \hat{z})$ for every latent element $\hat{y}_{i,c}$ at spatial position $i$ and channel $c$, where $\hat{y}_{<i,c}$ are the spatially preceding elements inside the same channel which are modelled by the masked context network  and  $\hat{y}_{<c}$ are the features derived from previously decoded channels at the same spatial position. 

The channel conditioning module processes the decoded feature maps of previous channels using a lightweight convolutional stack that captures statistical co-occurrence patterns among channels and concatenates them with the hyperprior features and the masked context features to form a unified conditioning vector. The combined feature vector is passed through the parameter estimation network to predict the conditional mean and scale parameters of the latent distribution. This hierarchical conditioning process ensures that the entropy model has access to all relevant contexts—global, spatial and inter-channel when estimating the probability of each latent element.

The conditioning is causal across channels. When encoding the $c$-th channel, only information from the first $(c-1)$ channels is used, maintaining the autoregressive decoding order. Due to the fact that cross-channel dependencies generally exhibit smoother and lower-frequency patterns compared to spatial dependencies, the channel conditioning module is relatively lightweight. It introduces minimal complexity compared to the spatial context model, while still providing a noticeable improvement in rate-distortion performance, particularly for natural images with significant inter-channel correlations in the latent domain. 

By conditioning on both spatially and inter-channel correlated components, the entropy model approximates the true posterior distribution of the latent variables more accurately:
\begin{equation}
p(\hat{y}|\hat{z}) = \prod_{c=1}^{C} \prod_i p(\hat{y}_{i,c}|\hat{y}_{<i,c}, \hat{y}_{<,c}, \hat{z})
\end{equation}

This model generalises upon the spatial autoregressive framework by expanding it to a two-dimensional dependency space that extends to both spatial and channel dimensions. In practice, this mechanism substantially lowers entropy residuals, resulting in increased compression efficiency while maintaining parallelism in training and stability during optimization.

\subsection{Slice Transform with Excitation}
Within the suggested method, the latent representation $\hat{y}$ is divided into channel slices that are decoded sequentially. Every slice goes through a dedicated slice transform that refines its latent features prior to parameter estimation and reconstruction. To improve this stage without considerable parameter overhead, the slice transform integrates a Squeeze-and-Excitation (SE) block that adaptively adjusts channel responses according to their overall contextual importance \cite{hu2018squeeze}. Convolutional feature maps often include redundancies or channels that contain uneven amounts of information. This discrepancy may lead to ineffective entropy modelling, as not all latent channels hold equal significance for the final reconstruction. The SE mechanism addresses this by learning channel-wise attention coefficients that adjust feature activations based on their statistical importance. This adaptive rescaling enables the network to concentrate its capacity on the more informative latent components, thus improving both the reconstruction quality and the rate–distortion performance.
Given a latent slice $Y \in \mathbb{R}^{H \times W \times C_s}$, where $H$ is the height of the slice, $W$ is the width and $C_s$ is the number of channels in it, the “squeeze” step summarises the global activation statistics of each channel into a descriptor through average pooling:
\begin{equation}
s = \frac{1}{H \cdot W} \sum_{i=1}^{H} \sum_{j=1}^{W} Y_{i,j}
\end{equation}

\begin{figure}[t]
\centering
\includegraphics[width=\columnwidth]{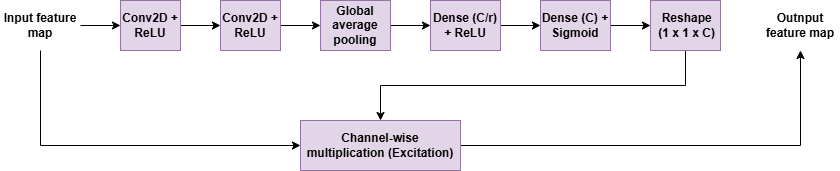}
\caption{The architecture of the squeeze-and-excitation block. C denotes the number of channels and r is the reduction ratio.}
\label{fig:seblock}
\end{figure}
The “excitation” step uses a small fully connected layer to capture nonlinear channel inter-dependencies:
\begin{equation}
w = \mathrm{sig}(W_2 \cdot \mathrm{relu}(W_1 \cdot s))
\end{equation}
where $sig$ is the sigmoid function, $relu$ is the rectified linear unit function (ReLU) and $W1, W2$ are learned weights. These weights serve as channel attention coefficients that rescale the feature responses via element-wise multiplication. This way, channels that contain more information for a given slice are amplified, while redundant or noisy ones are suppressed.

The SE block is incorporated into the slice transform, improving the transform's ability to capture subtle inter- channel dependencies and adjust representational focus dynamically across feature maps. In practice, this slice transformation with excitation results in a noticeable decrease in both entropy residuals and distortion after reconstruction. The architecture of the SE block is shown in Fig. \ref{fig:seblock}. 

\subsection{Latent Residual Prediction and Quantisation}
In neural image compression, rounding operations on the latent representation inevitably introduce quantization noise, which cannot be fully corrected by the hyperprior or masked context models. To overcome this limitation, the proposed framework incorporates a Latent Residual Prediction (LRP) module that estimates and compensates for residual errors in the latent domain after quantisation \cite{han2025causal}. 

The LRP module explicitly learns a correction term that refines each quantized latent slice by predicting the residual error conditioned on previously decoded features and hyperprior information. For the $m$-th latent slice, the refinement is expressed as:
\begin{equation}
\hat{y}'_m = \hat{y}_m + \lambda_{LRP} \cdot \text{softsign}(r_m)
\end{equation}
where $\hat{y}_m$ denotes the quantised slice, $r_m$ is the predicted residual and $\lambda_{LRP}$ is is a learned, trainable scaling factor to control the strength of the correction. Unlike previous approaches that employed the hyperbolic tangent, the proposed model adopts the softsign activation to ensure smooth gradients and bounded outputs, improving stability during training. 

Within this probabilistic model, the expected rate is defined as the cross-entropy between the learned prior and the real distribution of the quantised latents:
\begin{equation} \label{eq:rate2}
R = E_x \left[ -\log_2 \hat{p}_z (\hat{z}) - \sum_{c,i} \log_2 p(\hat{y}_{i,c} | \hat{y}_{<i,c}, \hat{y}_{<c}, \hat{z}) \right]
\end{equation}
The distortion measures the reconstruction error between the input and the decoded image:
\begin{equation}
    D = E_{x \sim p(x)} \left[ \| x - \hat{x} \|_2^2 \right]
\end{equation}
where $\hat{x} = g_s(\hat{y}';\theta_g)$ is the decoded image. 

\begin{table}[t]
\centering
\caption{Network Architecture}
\label{tab:network}
\resizebox{\columnwidth}{!}{%
\begin{tabular}{|l|l|l|}
\hline
\textbf{Component} & \textbf{Encoder} & \textbf{Decoder} \\
\hline
\textbf{Main Transform} & Conv 5$\times$5 c192 s2, GDN & Deconv 5$\times$5 c192 s2, IGDN \\
 & Conv 5$\times$5 c192 s2, GDN & Deconv 5$\times$5 c192 s2, IGDN \\
 & Conv 5$\times$5 c192 s2, GDN & Deconv 5$\times$5 c192 s2, IGDN \\
 & Conv 5$\times$5 cl\_d s2 & Deconv 5$\times$5 c3 s2 \\
\hline
\textbf{Hyper Transform} & Conv 3$\times$3 c320 s1, ReLU & Deconv 5$\times$5 c192 s2, ReLU \\
 & Conv 5$\times$5 c256 s2, ReLU & MaskConv 3$\times$3 c192 s1 \\
 & Conv 5$\times$5 ch\_d s2 & Deconv 5$\times$5 c256 s2, ReLU \\
 &  & MaskConv 3$\times$3 c256 s1 \\
 &  & Deconv 3$\times$3 c320 s1 \\
\hline
\textbf{Slice Transform} & \multicolumn{2}{c|}{Conv 5$\times$5 c224 s1, ReLU + SE Block} \\
 & \multicolumn{2}{c|}{Conv 5$\times$5 c128 s1, ReLU} \\
 & \multicolumn{2}{c|}{Conv 3$\times$3 cs\_d s1} \\
\hline
\multicolumn{3}{p{1.0\linewidth}}{\footnotesize
Conv: convolutional layer, Deconv: deconvolutional layer,
c: channels, s: stride, l\_d, latent depth,
MaskConv: masked convolution, h\_d: hyperprior depth,
s\_d: latent depth / number of slices
}
\end{tabular}% 
}
\end{table}

\begin{figure*}[t]
\centering
\includegraphics[width=0.9\textwidth]{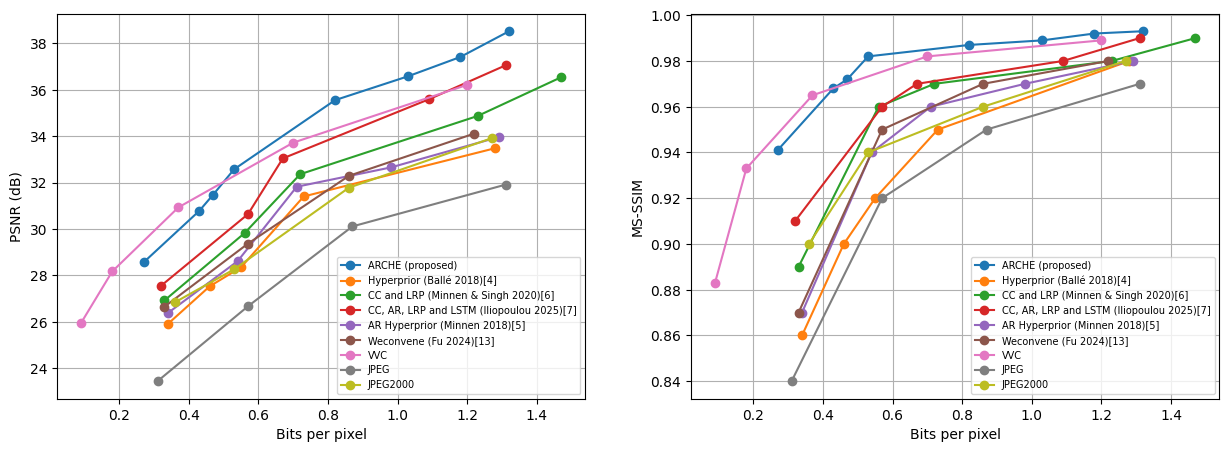}
\caption{Rate-distortion curves on Kodak dataset. PSNR (left) and MS-SSIM (right) versus bit rate. ARCHE achieves superior performance across most operating points, surpassing VVC intra.}
\label{fig:rd_curves}
\end{figure*}

\begin{figure*}[t]
\centering
\includegraphics[width=0.9\textwidth]{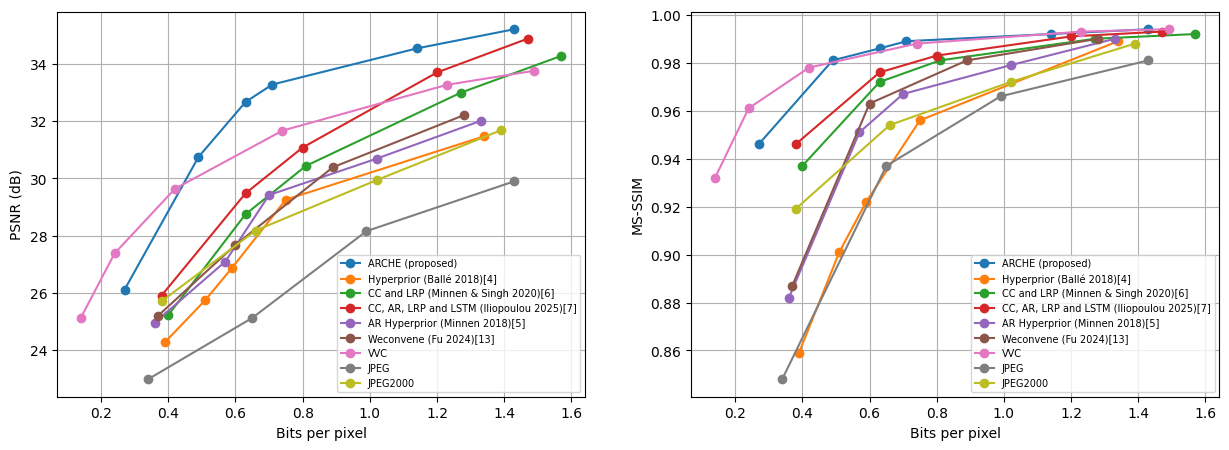}
\caption{Rate-distortion curves on Tecnick dataset. PSNR (left) and MS-SSIM (right) versus bit rate. ARCHE achieves superior performance across most operating points, surpassing VVC intra.}
\label{fig:rd_curves2}
\end{figure*}

\section{Experimental Setup}
This section details the implementation and training setup of the proposed framework, along with the evaluation benchmarks and baseline methods against which it is compared. 

All the models were trained on the CLIC set, which contains high-resolution natural images \cite{clic2024}. The dataset was converted to PNG format and loaded through a TensorFlow data pipeline with random shuffling. During training, random 256×256 crops were extracted from each image and normalised to the [0,1] range. The evaluation was performed on a widely used benchmark set, Kodak, that consists of 24 natural images sized 768$\times$512 \cite{kodak1993}. This dataset was selected due to its wide range of textures, edges and lighting conditions, making it suitable for measuring generalisation in learned compression. During inference, no data augmentation or resizing was used, and all metrics were calculated in the RGB domain with 8-bit precision. All models were trained and evaluated under identical conditions to ensure a fair comparison.

The proposed model follows the architecture described in Section III, implemented entirely in TensorFlow 2.11 using the TensorFlow Compression (TFC) library. The layers of the compression network are presented in Table \ref{tab:network}. The latent depth was fixed to 320, divided into 10 slices for progressive entropy decoding. Each slice employs its own conditional transform with SE-based excitation and a separate LRP sub-module. The hyperprior depth was set to 192 and the reduction ratio in the SE block was 16. 

All models were trained end-to-end to minimize the rate–distortion loss described in (1). Separate models were trained for eight values of the trade-off parameter $\lambda \in \{0.001, 0.005, 0.007, 0.01, 0.03, 0.05, 0.07, 0.1\}$ covering the operating range from nearly lossless to highly compressed conditions. For the experimental results, the model was trained for 400 epochs with batch size 8, using the Adam optimizer with a learning rate of $10^{-4}$. The distortion metric that was used for the optimisation was the mean squared error (MSE). The code is available on https://github.com/sof-il/ARCHE. Both the training and the testing of the model were implemented on an NVIDIA GeForce RTX 3080, 11GB. 

The proposed approach was compared against six representative image compression baselines, both traditional and learning based:
\begin{itemize}
\item JPEG, a widely used transform-coding standard, based on the discrete cosine transform (DCT).

\item JPEG2000, a more advanced traditional standard, employing the discrete wavelet transform (DWT).

\item VVC Intra, the latest video compression standard that also performs intra-frame coding \cite{vvc2020}.

\item Ballé et al., the original scaled hyperprior model \cite{balle2018variational}.

\item Minnen et al., joint autoregressive and hierarchical priors \cite{minnen2018joint}.

\item Minnen \& Singh, a channel-wise autoregressive entropy model \cite{minnen2020channelwise}.

\item WeConvene, a wavelet-based learned codec \cite{fu2024weconvene}.

\item Iliopoulou et al., the authors’ earlier framework that incorporates latent residual prediction and channel conditioning but does not include excitation and uses an LSTM-based context model \cite{iliopoulou2025learned}.
\end{itemize}

\section{Results and Discussion}
In this section, the performance of the proposed model is compared to the image compression techniques mentioned in Section IV. Additionally, an ablation study is added to evaluate the contribution of each component of the proposed framework as well as investigate the effect of the number of channel slices on its performance.

\subsection{Rate-Distortion Performance}
\begin{table}[b]
\centering
\caption{BD-Rate Savings (\%) vs Ballé \cite{balle2018variational} and VVC Intra on Kodak (PSNR)}
\label{tab:bdrate}
\begin{tabular}{|l|c|c|}
\hline
\textbf{Method} & \textbf{BD-Rate vs Ballé} & \textbf{BD-Rate vs VVC} \\
\hline
Minnen et al. \cite{minnen2018joint} & -8.00 & 90.61 \\
\hline
Minnen \& Singh \cite{minnen2020channelwise} & -16.28 & 63.55 \\
\hline
WeConvene \cite{fu2024weconvene} & -6.92 & 92.47 \\
\hline
Iliopoulou et al. \cite{iliopoulou2025learned} & -24.22 & 30.19 \\
\hline
\textbf{ARCHE (proposed)} & \textbf{-48.01} & \textbf{-5.61} \\
\hline
\end{tabular}
\end{table}
The rate–distortion (R-D) performance of ARCHE is evaluated on the Kodak benchmark and compared against the codecs described in Section IV. ARCHE consistently outperforms all baselines across most bit rates, including the traditional VVC intra codec. Fig. \ref{fig:rd_curves} and \ref{fig:rd_curves2} show rate-distortion curves on the Kodak and Tecnick datasets respectively. BD-rate savings relative to Ballé \cite{balle2018variational} and VVC Intra are presented in Tables \ref{tab:bdrate} and \ref{tab:bdrate2}, demonstrating that ARCHE achieves substantial gains over both traditional and learned methods. The distortion metrics that are used in this paper are the PSNR and the MS-SSIM.

Across most operating points, the proposed model consistently achieves lower bitrates for the same reconstruction quality, demonstrating improved coding efficiency. Compared to the hyperprior, ARCHE provides a substantial bit rate reduction while retaining comparable perceptual quality. Compared to the joint autoregressive and hyperprior model, the channel-wise autoregressive model, and WeConvene, the proposed method also gives superior results, proving the effectiveness of the added components in image compression. The improvement is due to the combined effect of masked spatial context modelling, channel conditioning, and the latent residual prediction mechanism, which together produce more accurate conditional probability estimates in the entropy model. ARCHE displays smaller but still noticeable improvement in the compression results relative to the \cite{iliopoulou2025learned} framework. This is attributed to the substitution of a masked context model for an LSTM-based one, as well as the addition of excitation in the slice transform. Similar patterns emerge in the assessment of MS-SSIM, with ARCHE demonstrating enhanced performance in the mid-to-low-bit rate range, producing visually clearer reconstructions with fewer texture artifacts. Overall, the model achieves state-of-the-art compression efficiency among methods of similar architectural complexity.
\begin{table}[t]
\centering
\caption{BD-Rate Savings (\%) vs Ballé \cite{balle2018variational} and VVC Intra on Tecnick (PSNR)}
\label{tab:bdrate2}
\begin{tabular}{|l|c|c|}
\hline
\textbf{Method} & \textbf{BD-Rate vs Ballé} & \textbf{BD-Rate vs VVC} \\
\hline
Minnen et al. \cite{minnen2018joint} & -8.81 & 79.04 \\
\hline
Minnen \& Singh \cite{minnen2020channelwise} & -13.99 & 50.32 \\
\hline
WeConvene \cite{fu2024weconvene} & -6.77 & 86.01 \\
\hline
Iliopoulou et al. \cite{iliopoulou2025learned} & -22.71 & 33.59 \\
\hline
\textbf{ARCHE (proposed)} & \textbf{-44.89} & \textbf{-10.28} \\
\hline
\end{tabular}
\end{table}

\begin{figure}[t]
\centering
\includegraphics[width=0.9\columnwidth]{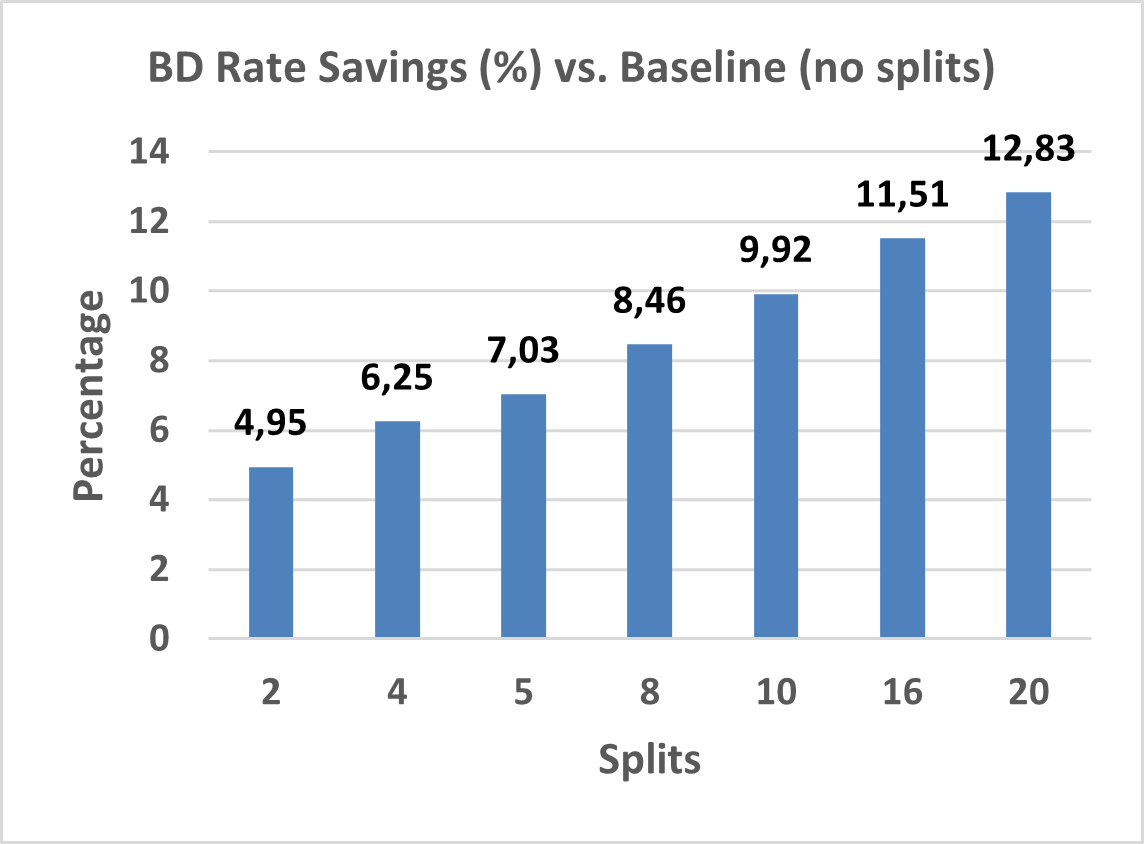}
\caption{BD-rate savings versus number of channel slices. Performance saturates around 10 slices.}
\label{fig:slices}
\end{figure}

\begin{figure}[t]
\centering
\includegraphics[width=0.9\columnwidth]{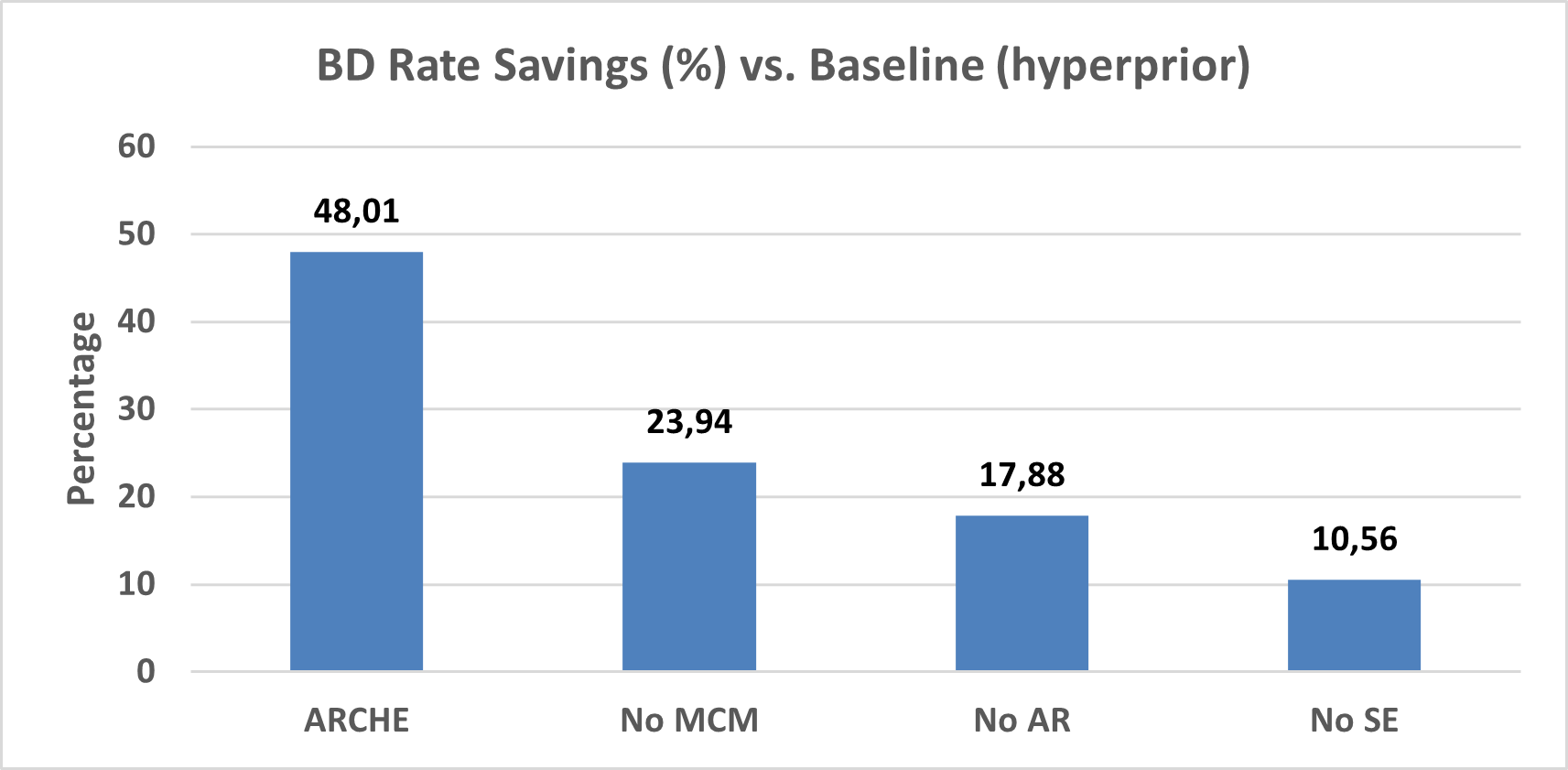}
\caption{BD-rate savings of component ablations versus hyperprior baseline. Each module contributes complementary improvements.}
\label{fig:ablation}
\end{figure}

\subsection{Ablation Study}
We evaluated ARCHE with varying numbers of latent slices while keeping all other architectural and training settings identical, in order to assess the impact of latent partitioning on compression performance. Every configuration progressively increases the detail of the latent representation, allowing better modelling of channel dependencies and improving entropy estimation accuracy. Fig. \ref{fig:slices} illustrates the resulting BD-rate savings compared to the baseline configuration that uses a single latent slice.

The results show a consistent improvement in performance as the number of slices increases. Using only two slices already 
yields a notable gain of nearly 5\%, while configurations with 10 or more slices achieve over 11\% average BD-rate savings. This trend indicates that splitting the latent space into smaller, sequential groups enables the entropy model to better exploit inter-slice dependencies and statistical correlations. However, the incremental benefit gradually decreases, suggesting that excessive slicing offers limited additional compression efficiency while increasing computational overhead. For this reason, a slice count of 10 was selected for the final architecture to balance coding efficiency with inference cost.

To evaluate the individual impact of each architectural component, we conducted a systematic ablation study in which individual modules were removed or combined. The evaluated model variants include: (i) without the masked context model (no MCM), (ii) without the context model and the autoregressive prior (no AR), and (iii) without squeeze-and-excitation in the slice transform (no SE). Fig. \ref{fig:ablation} presents the resulting BD-Rate savings over the hyperprior baseline.

Eliminating both spatially autoregressive components yields the most significant drop in the rate–distortion performance, since it reduces the system to a hyperprior-only model, substantially worsening the rate-distortion performance by eliminating all spatial dependency modelling, but allowing fully parallel decoding with lower computational cost. Removing the masked context model also significantly reduces the compression results, confirming that spatially masked context modelling is essential for precise local probability estimation. Finally, excluding the excitation mechanism results in a moderate decline in quality at lower bit rates, suggesting that adaptive channel re-weighting is essential for preserving fine-grained structural details. Overall, these results demonstrate that each component contributes a distinct benefit, and their combination achieves the most consistent rate-distortion improvement across all bit rates.

\subsection{Visual Comparisons}
\begin{figure}[t]
\centering
\includegraphics[width=\columnwidth]{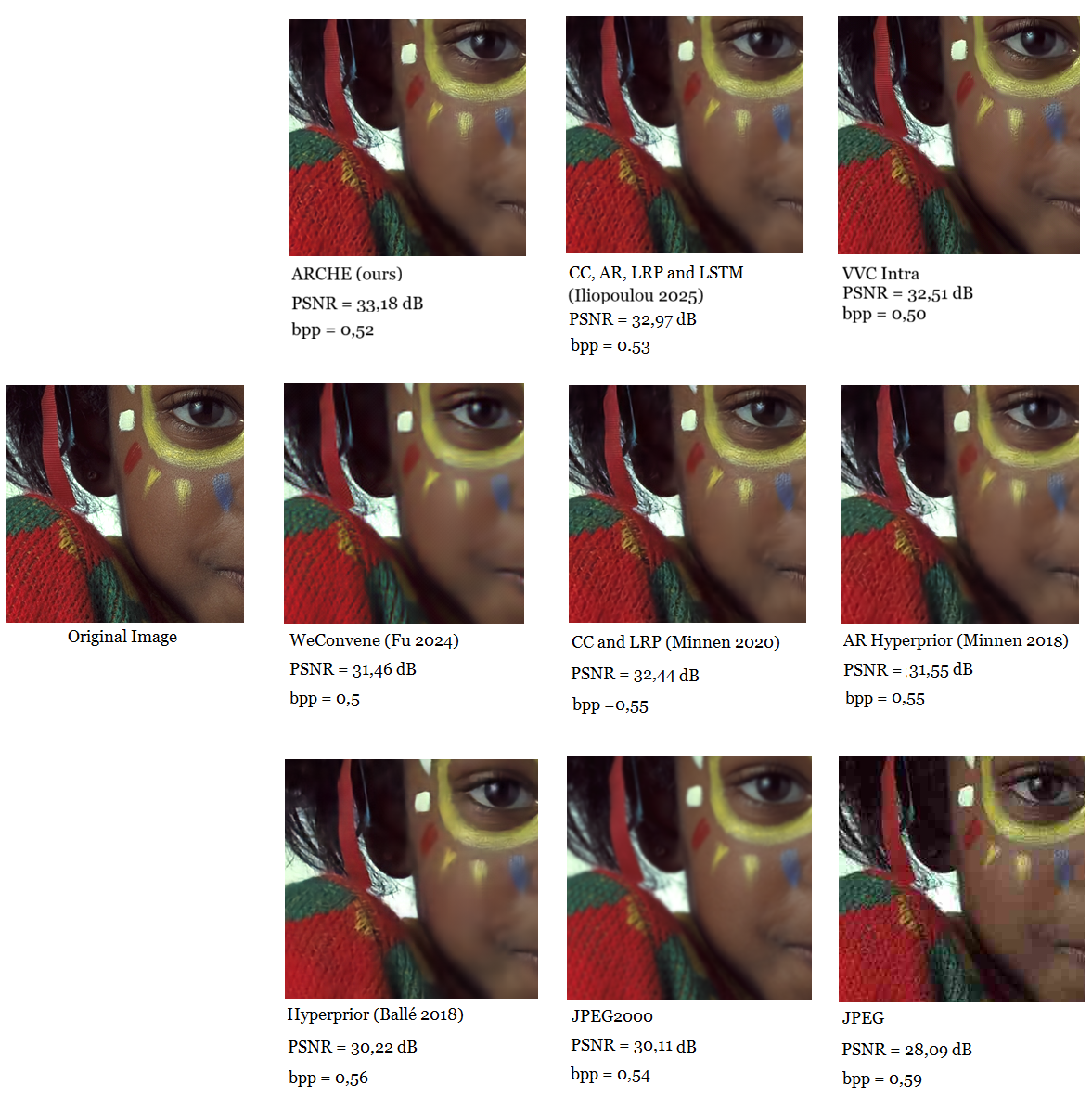}
\caption{Visual comparison on Kodak at similar bit rates. ARCHE preserves sharper details and more natural colours.}
\label{fig:visual}
\end{figure}

Visual comparisons between the proposed method and other compression techniques are presented in Fig. \ref{fig:visual}, highlighting the superior reconstruction quality achieved by our approach. ARCHE generates reconstructed images with sharper edges, better texture detail, and more accurate preservation of colour gradients. The differences are most noticeable in areas containing complex spatial structures such as foliage, fabric, or small-scale patterns, where other models tend to introduce blurring or ringing artifacts. Visually, the proposed approach yields more natural colour transitions and fewer high- frequency distortions, consistent with the results presented in the rate-distortion analysis.

\subsection{Computational Complexity}
Beyond coding performance, an important aspect of learned image compression is computational efficiency. Therefore, this subsection examines the complexity of the proposed architecture and its decoding speed during inference. The proposed architecture contains approximately 95 million parameters, comparable to existing autoregressive methods but substantially lighter than transformer-based variants. Measured on an NVIDIA GeForce RTX 3080 GPU, the average decoding time per Kodak image is 222 ms. Therefore, the computational cost of ARCHE is similar to other similar methods, despite having the addition of excitation. Table \ref{tab:complexity} summarises computational metrics.
\begin{table}[t]
\centering
\caption{Model Complexity and Inference Time per image}
\label{tab:complexity}
\begin{tabular}{|l|c|c|}
\hline
\textbf{Method} & \textbf{Parameters} & \textbf{Time (ms)} \\
\hline
Ballé et al. \cite{balle2018variational} & 11.7M & 25 \\
\hline
Minnen et al. \cite{minnen2018joint} & 95.8M & 591 \\
\hline
Minnen \& Singh \cite{minnen2020channelwise} & 121.7M & 249 \\
\hline
Iliopoulou et al. \cite{iliopoulou2025learned} & 124.3M & 265 \\
\hline
\textbf{ARCHE (proposed)} & \textbf{95.4M} & \textbf{222} \\
\hline
\end{tabular}
\end{table}

Autoregressive entropy models based on ConvLSTMs capture spatial dependencies by maintaining a recurrent state that propagates across rows and columns of the latent feature map \cite{johnston2018improved,iliopoulou2025learned}. While this approach is powerful, it is also quite computationally expensive. Each latent element must be decoded in order, as its probability depends on all previously decoded states, resulting in serial decoding. This significantly slows entropy decoding, as arithmetic coding must also proceed sequentially. ConvLSTMs require maintaining hidden and cell states over large areas, resulting in substantial memory consumption and increased latency, particularly for high-resolution images. Finally, training recurrent entropy models can be less stable due to long-range temporal dependencies and vanishing gradient effects. 

On the other hand, Masked PixelCNN models maintain causality directly through masked convolutions rather than through explicit recurrence. This design enables parallel probability estimation within each partially decoded context element, improving training stability and greatly reducing computational cost.

\section{Conclusion}
This paper introduces ARCHE – Autoregressive Residual Compression with Hyperprior and Excitation, a unified learned image compression framework that balances expressiveness, efficiency, and practicality. By incorporating hierarchical, spatial, and channel-wise priors into a single probabilistic architecture, ARCHE captures both global and local dependencies in the latent space without relying on complex and computationally expensive recurrent or transformer-based components. The model achieves significant improvements in rate–distortion efficiency, showing reductions of approximately 48\% BD-rate relative to the widely used hyperprior baseline of Ballé et al. and 30\% compared to the channel-wise autoregressive model of Minnen \& Singh on the Kodak dataset, while remaining computationally lightweight with 95M parameters and 222 ms per image decoding time.

The results demonstrate that the careful design of the system architecture can offer an interesting alternative to the trend of increasingly larger and slower models. Instead of seeking higher complexity, ARCHE improves the compression performance by deepening the interaction among entropy estimation, contextual dependency capture, and adaptive feature recalibration, since these modules are complementary to each other. Each of the proposed modules plays a role in this. The masked context model refines spatial dependency modelling, the channel conditioning enables inter-channel modelling without excessive overhead, the excitation mechanisms dynamically reweight latent responses to emphasize features with more information and the latent residual prediction gives a focused correction for quantisation artifacts. Their combination produces a cohesive architecture that is capable of achieving near state-of-the-art visual quality while maintaining the efficiency and robustness of fully convolutional architectures. 

Apart from the numerical results, the qualitative findings of the model are just as important. Visual comparisons across the Kodak dataset benchmark show that ARCHE consistently reconstructs images with sharper edges and more natural colour transitions, especially in areas with fine texture where other techniques tend to blur or over-smooth details. This balance between quantitative performance and perceptual accuracy highlights the potential of probabilistic convolutional models as practical alternatives to heavier attention-based architectures, especially in cases where speed and computational efficiency are essential.

The ablation studies further highlight several valuable observations about the importance of design in learned compression. Increasing the number of channel slices improves the compression performance by allowing finer-grained conditional modelling between latent groups, but with decreasing gains after a certain number of splits. This result suggests a trade-off between entropy modelling detail and computational cost, offering a guideline for designing architectures tailored to specific deployment constraints. Likewise, the study of individual components confirms that each module – masked context, autoregression and excitation – contributes a distinct and complementary advantage to the performance of the system. These findings reinforce the importance of structured hybrid designs that unite hierarchical priors, context adaptation, and lightweight attention instead of relying on a primary mechanism.

While the results are encouraging, several directions remain open for exploration. One promising path involves further improving parallel processing during entropy modelling. Although masked convolutional models are much faster than recurrent ones (e.g., LSTM-based ones), their design is strictly causal, which imposes some degree of sequential processing. Future work could explore semi-parallel decoding strategies, such as block-wise context prediction, to accelerate inference without sacrificing rate-distortion performance. Another possible direction lies in perceptual and task-oriented optimization. The present work optimises for mean squared error. However, incorporating perceptual metrics could further enhance the visual realism of the reconstructed images, especially at low bit rates. Additionally, ensemble learning approaches could be explored to improve rate-distortion optimization robustness across diverse content types, as demonstrated by Wang et al. \cite{wang2021ensemble}, which may offer complementary benefits when combined with the hierarchical entropy modelling framework proposed here. Similarly, compression can be tailored for certain vision tasks like classification or segmentation, resulting in joint representations that serve both efficient compression and analysis. Recent work by Fischer et al. has shown that latent space masking techniques can optimise compression for machine vision tasks by selectively preserving task-relevant features \cite{fischer2025boosting}, suggesting a promising direction for extending ARCHE beyond human perception. This way, ARCHE could evolve into a general-purpose latent modelling framework that applies to more than image reconstruction.

In summary, ARCHE contributes to the constant evolution of learned image compression by showing that architectural consistency and targeted efficiency improvements can rival the performance of more complex models. By reexamining the balance between statistical expressiveness and computational applicability, this study highlights that achieving high compression efficiency does not necessarily demand increased architectural depth but rather a better understanding of dependency modelling. Future work will continue along this trajectory – refining, generalizing, and utilising such models toward scalable and adaptive compression systems.

\clearpage
\appendices
\section{Gaussian Mixture Entropy Model}\label{appendix:A}

The baseline approach models each latent element with a single Gaussian distribution, which is a simplified approximation of its true latent distribution. In practice, latent representations of natural images often display a slight multimodality, as a result of the nonlinear effects of the transform and context networks. To investigate whether a richer probabilistic model can further improve rate–distortion performance, we extend the entropy model to a Gaussian mixture formulation.

In the baseline model, each latent coefficient $\hat{y}_{i,c}$ is modelled by a single Gaussian distribution parameterized by a mean and scale, conditioned on the hyperprior, masked spatial context, and channel dependencies:
\begin{equation}
p(\hat{y}_{i,c} \mid \Psi) = \mathcal{N}(\hat{y}_{i,c}; \mu_{i,c}, \sigma_{i,c}^2)
\end{equation}
where $\Psi$ are the total conditioning features mentioned in the previous sections. To address the restrictions of this model, we extend the entropy model to a Gaussian Mixture Model (GMM), defined as:
\begin{equation}
(\hat{y}_{i,c} \mid \Psi) = \sum_{k=1}^{K} w_{i,c,k} \mathcal{N}(\hat{y}_{i,c}; \mu_{i,c,k}, \sigma_{i,c,k}^2)
\end{equation}
where $K$ is the number of Gaussian distributions used and  $w_{i,c,k}$ are the normalized mixture weights that satisfy two mathematical relations: (i) $w_{i,c,k}>0,\forall k$ and (ii) $\sum_{k} w_{i,c,k} = 1$. The parameters $\mu_{i,c,k}$, $\sigma ^2_{i,c,k}$, and $w_{i,c,k}$ are not learned globally but are predicted adaptively by a neural network. A lightweight $1 \times 1$ convolutional layer outputs $3 \cdot K$ channels per latent element, corresponding to the means, scales and logits for each mixture component and spatial position, using as input the conditioning features derived from the hyperprior, masked context model and previously decoded latent slices. Since neural networks cannot directly produce positive mixture weights that sum to one, the network predicts unnormalized logits—real-valued outputs that are transformed via a log-sum-exp (softmax) normalization to yield valid mixture weights. The expected rate is defined as:
\begin{equation}
R = -\frac{1}{M} \sum_{i,c} \log_{2} \left( \sum_{k=1}^{K} w_{i,c,k} \mathcal{N}(\hat{y}_{i,c}; \mu_{i,c,k}, \sigma_{i,c,k}^2) \right)
\end{equation}
\begin{figure}[t]
\centering
\includegraphics[width=0.8\columnwidth]{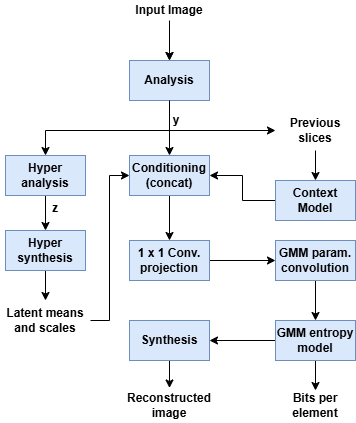}
\caption{The flow chart of the incorporation of the GMM into the ARCHE pipeline.}
\label{fig:gmm}
\end{figure}

where $M$ denotes the number of latent elements. The GMM parameters are optimized jointly with all other components using the same rate–distortion loss as in \eqref{eq:rate}. The Gaussian mixture variant introduces only minor additional computational overhead, as the extra convolutional layers predicting the mixture parameters are lightweight. Despite this, the GMM provides improved flexibility in representing   multimodal latent distributions, leading to smoother rate–distortion curves and more accurate probability estimation at low bitrates. Fig. \ref{fig:gmm} presents a flow of the integration of the GMM into the ARCHE architecture. 

\section{Checkerboard Context Model}
The masked PixelCNN-based context model employed in ARCHE captures fine-grained spatial dependencies through causal masked convolutions, providing accurate local probability estimates. However, it processes latent elements in strictly sequential decoding order (one at a time in raster-scan fashion), which slows down the decoding process. In order to investigate the trade-off between decoding efficiency and rate–distortion performance, we substitute the masked PixelCNN context model with a checkerboard context model (CCM), as originally proposed by He et al. \cite{he2021checkerboard}.

\begin{figure}[t]
\centering
\includegraphics[width=0.9\columnwidth]{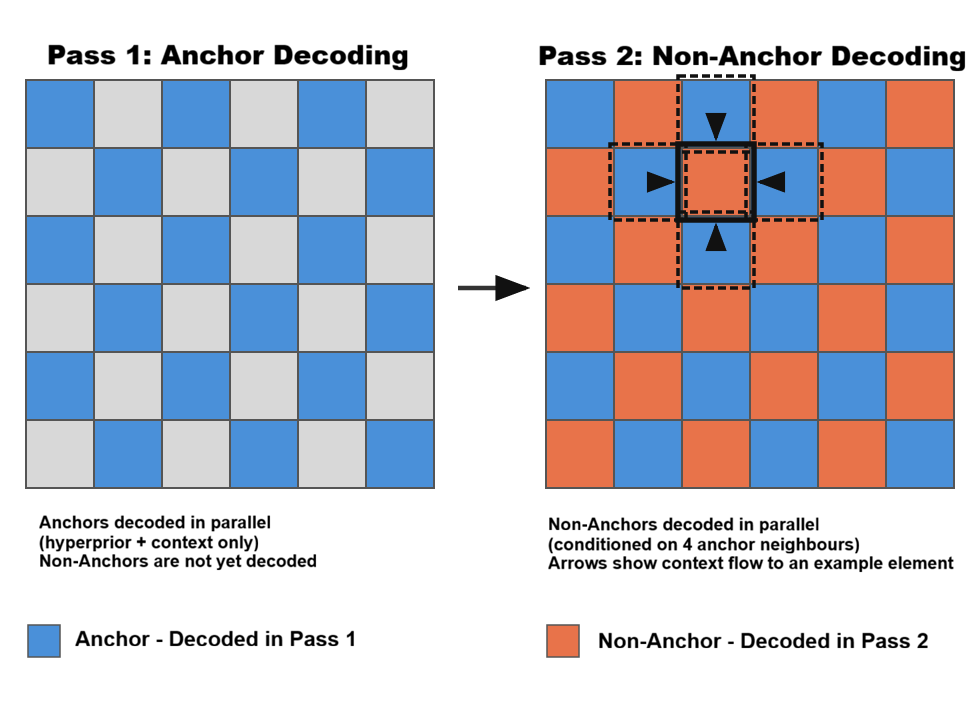}
\caption{Two-pass checkerboard decoding. The arrows show the four anchor neighbours giving spatial context to each non-anchor element.}
\label{fig:checkerboard}
\end{figure}

The checkerboard context model modifies the spatial decoding order to allow latent elements to be processed in two parallel passes instead of sequentially. The latent feature map $hat{y}$ is divided into two interleaved sets based on the parity of the spatial index sum $(i+j)$. Elements with even values of $(i+j)$ are treated as anchor elements, while those with odd values are non-anchor elements, forming a checkerboard-like spatial pattern as shown in Fig. \ref{fig:checkerboard}.

In the first pass, all anchor elements are decoded simultaneously without using any spatial context. Their probability distributions are conditioned only on the hyperprior, channel conditioning from previously decoded slices, and the features produced by the SE-based slice transform.The conditional distribution of each anchor element is given by:
\begin{equation}
p(\hat{y}_{i,j,c} \mid \hat{y}_{<c}, \hat{z}) = \mathcal{N}(\hat{y}_{i,j,c}; \mu_{i,j,c}, \sigma_{i,j,c}^2), \quad (i+j) \text{ even}
\end{equation}

In the second pass,  all non-anchor elements are decoded simultaneously, each conditioned on its four spatially adjacent anchor neighbours (above, below, left, and right) which are already available from the first pass. A masked convolution with a $3 \times 3$ or $5 \times 5$ checkerboard kernel is applied to extract the local context features from these neighbours. The conditional distribution for non-anchor elements is:
\begin{equation}
p(\hat{y}_{i,j,c} \mid \hat{y}_{N(i,j),c}, \hat{y}_{<c}, \hat{z}) = \mathcal{N}(\hat{y}_{i,j,c}; \mu_{i,j,c}, \sigma_{i,j,c}^2), \quad (i+j) \text{ odd}
\end{equation}
where $\hat{y}_{N(i,j),c}$ denotes the set of already-decoded anchor neighbours at positions adjacent to $(i,j)$ in the same channel slice $c$.

The context features extracted in the second pass are concatenated with the hyperprior and channel conditioning features and processed by the same parameter estimation network used in the main ARCHE architecture. This network predicts the conditional Gaussian parameters $(\mu, \sigma)$ for each latent element. The resulting joint distribution over the latent tensor factorises as:
\begin{equation}
\begin{split}
p(\hat{y} \mid \hat{z}) = \prod_{c=1}^{C} \Biggl( & \prod_{i,j : (i+j) \text{ even}} p(\hat{y}_{i,j,c} \mid \hat{y}_{<c}, \hat{z}) \\
& \cdot \prod_{i,j : (i+j) \text{ odd}} p(\hat{y}_{i,j,c} \mid \hat{y}_{N(i,j),c}, \hat{y}_{<c}, \hat{z}) \Biggr)
\end{split}
\end{equation}
Compared to the masked PixelCNN context model, the CCM reduces the number of sequential decoding steps from $H \times W$ per channel slice to just two, resulting in significant improvement in decoding speed. However, this comes at the cost of a small reduction in the amount of spatial context available to anchor elements.

In addition to substituting the masked PixelCNN with the checkerboard context model, this variant introduces a modification to the latent slicing process. Rather than dividing the 320 latent channels uniformly across 10 slices, a content-adaptive slicing strategy is adopted, assigning the following channel depths to slices 1 through 10: [48,40,36,32,32,32,28,24,24,24]. This non-uniform allocation is motivated by the asymmetry in the conditioning context available to each slice. Early slices are decoded with access to fewer previously decoded channels and, in the checkerboard model, with no spatial context for anchor elements in the first pass. As a result, these slices require greater representational capacity to model their latent distributions accurately. In contrast, later slices benefit from richer channel conditioning accumulated from previous slices, allowing their distributions to be captured with fewer parameters. By allocating more capacity to the earlier, less-conditioned slices, this adaptive slicing partially compensates for the reduced spatial context of the checkerboard design and improves entropy modelling accuracy where it is most needed.

Apart from the context modelling and adaptive slicing strategy, this variant remains architecturally identical to ARCHE. It utilises the same hyperprior, SE-based slice transforms and latent residual prediction modules. The entropy model is still trained using the same rate–distortion objective defined in \eqref{eq:rate} and \eqref{eq:rate2} of the main paper. The experimental results comparing the three context model variants are presented in the following section.

\begin{figure*}[t]
\centering
\includegraphics[width=0.9\textwidth]{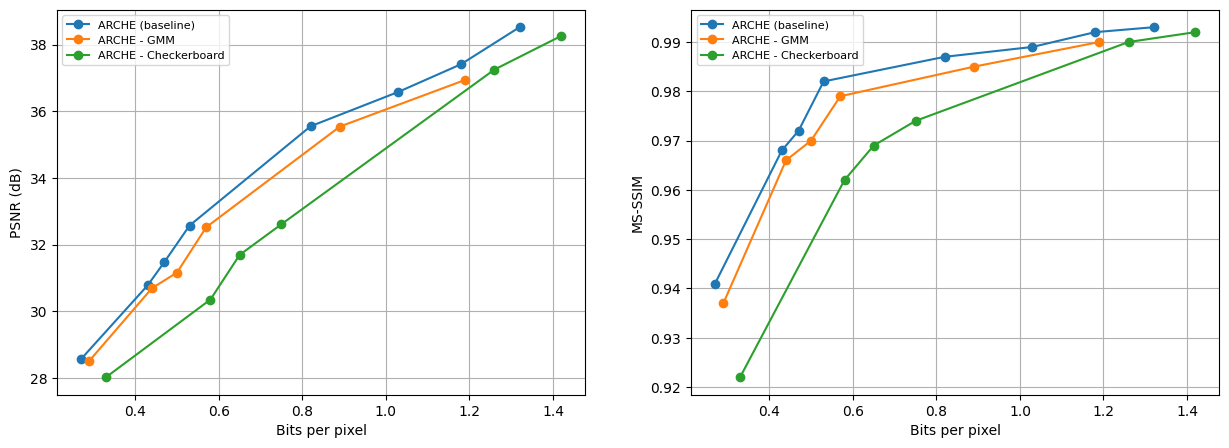}
\caption{Rate-distortion curves on Kodak dataset. PSNR (left) and MS-SSIM (right) versus bit rate. ARCHE achieves superior performance across all operating points.}
\label{fig:rd_gmm_kodak}
\end{figure*}

\begin{figure*}[t]
\centering
\includegraphics[width=0.9\textwidth]{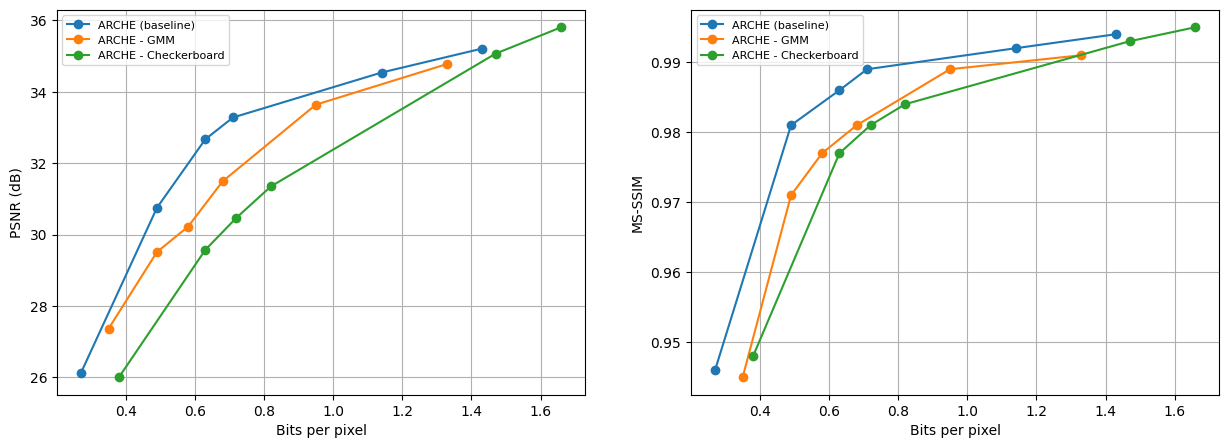}
\caption{Rate-distortion curves on Tecnick dataset. PSNR (left) and MS-SSIM (right) versus bit rate. ARCHE achieves superior performance across all operating points.}
\label{fig:rd_gmm_tecnick}
\end{figure*}

\section{Results Comparison}
In this section, the performance of the proposed variants is
compared to that of the ARCHE baseline framework. Fig. \ref{fig:rd_gmm_kodak} and \ref{fig:rd_gmm_tecnick} present the rate–distortion curves on the Kodak benchmark.

Although mixture-based priors have been shown to improve modelling flexibility in some learned compression frameworks, in our case the results did not indicate a significant improvement in rate–distortion performance after replacing the single-Gaussian entropy model with a Gaussian Mixture likelihood. The GMM-based variant closely follows the baseline ARCHE curve, with marginal or no gains across the tested bit rate range. This indicates that the hierarchical and contextual components of ARCHE already provide sufficient probabilistic modelling, leaving little residual correlation for the GMM to exploit. Moreover, the additional mixture parameters slightly increase the model complexity and training time without yielding considerable benefits in reconstruction quality. Overall, these findings indicate that the single-Gaussian entropy model offers an adequate trade-off between expressiveness and efficiency within the ARCHE framework, while more complex priors such as Gaussian Mixtures may only provide significant improvements in settings with weaker context modelling.

Substituting the masked PixelCNN context model with the checkerboard context model results in a more pronounced rate–distortion degradation, particularly at low bit rates. This behaviour is consistent with the fundamental design trade-off of the checkerboard model: by restricting anchor elements to hyperprior and channel context only,  without any spatial neighbourhood information in the first pass, approximately half of all latent elements are encoded with weaker probability estimates. The non-anchor elements benefit from four anchor neighbours in the second pass, but this asymmetry in context availability limits the overall entropy modelling accuracy compared to the full raster-scan spatial autoregression of the masked PixelCNN. The narrowing of the performance gap at high bit rates is expected, as distortion becomes the dominant term in the rate–distortion objective and the contribution of entropy modelling precision lessens. 

Despite the rate–distortion cost, the checkerboard variant offers significant gains in training efficiency. In practice, the checkerboard variant achieved a 58\% reduction in training time compared to the ARCHE baseline, as the two-pass parallel processing allows gradient computation to proceed across all anchor and non-anchor elements simultaneously within each pass, avoiding the element-wise sequential dependencies of the masked PixelCNN during backpropagation. However, contrary to the expected improvement in decoding parallelism, the checkerboard variant was approximately 15\% slower than the ARCHE baseline during inference. This behaviour is most likely caused by the overhead associated with the two-pass decoding process, which includes intermediate buffering of anchor states and the synchronisation required between decoding stages as well as the variable-width slice processing introduced by the content-adaptive slicing scheme, which may interact less efficiently with the fixed memory layout of the parameter estimation network at the spatial resolutions used in our experiments. These additional operations appear to outweigh the benefits of increased parallelism. The results indicate that the theoretical inference speedup of the checkerboard design may only become evident at much higher image resolutions, where the sequential raster-scan decoding of the masked PixelCNN model becomes the dominant computational bottleneck. Within the limitations of the present work, the checkerboard variant is therefore more suitable as a training-efficient alternative than as a deployment-oriented codec.

Taken together, these experiments highlight two potential extensions of ARCHE that ultimately do not improve upon the baseline. The GMM variant introduces additional parametric flexibility into the entropy model but does not provide any measurable rate–distortion benefit. This suggests that a single Gaussian prior is already well aligned with the latent statistics produced by the full conditioning pipeline. On the other hand, the checkerboard variant leads to a noticeable degradation in rate–distortion performance and, within the resolution range of the Kodak and Tecnick datasets, does not provide the inference speedup that motivates its design. However, its significantly shorter training time may still make it a practical option in scenarios that require rapid experimentation or frequent retraining. For these reasons, neither variant is incorporated into the final ARCHE architecture. Nevertheless, both modifications remain fully compatible with the rest of the framework, and their behaviour under different image resolutions or alternative training remains a possible direction for future work.
\end{document}